\documentclass[aps,prb,showpacs,twocolumn,amsmath,amssymb,superscriptaddress]{revtex4-1}
\usepackage[usenames,dvipsnames]{xcolor}
\usepackage{float}
\usepackage{amsfonts}
\usepackage[english]{babel}
\usepackage[T1]{fontenc}
\usepackage{times}
\usepackage{mathrsfs}
\usepackage{graphicx}
\usepackage{bm}
\renewcommand{\vec}[1]{\boldsymbol{#1}}

\begin{document}

\title{Disorder Effects on
Helical Edge Transport in Graphene under a Strong, Tilted Magnetic Field}
\author{Chunli Huang}

\affiliation{Division of Physics and Applied Physics, School of Physical and Mathematical Sciences, Nanyang
Technological University, Singapore 637371, Singapore}
\affiliation{Department of Physics, National Tsing Hua University, Hsinchu 30013, Taiwan}
\author{Miguel A. Cazalilla}
\affiliation{Department of Physics, National Tsing Hua University, Hsinchu 30013, Taiwan}
\affiliation{National Center for Theoretical Sciences (NCTS), National Tsing Hua University, Hsinchu 30013, Taiwan}
\affiliation{Donostia Internatinal Physics Center (DIPC), Manuel de Lardizabal, 4. E-20018 San Sebastian, Spain.}
\date{\today}

\begin{abstract}
 In a recent experiment, Young et al. [Nature {\bf 505}, 528 (2014)] observed a metal to insulator transition as well as transport through helical edge states in monolayer graphene under a strong, tilted magnetic field. Under such conditions, the bulk is a magnetic insulator which can exhibit metallic conduction through helical edges.  It was found that the two-terminal conductance of the helical channels  deviates from the expected quantized value ($=e^2/h$ per edge, at zero temperature). Motivated by this observation, we study the effect of 
disorder on the conduction through the edge channels. We show that, unlike for helical edges of topological insulators in semiconducting quantum wells, a disorder Rashba spin-orbit coupling does not lead to backscattering, at least to leading order. Instead, we find that the lack of perfect anti-alignment of the electron spins in the helical  channels  to be the most likely cause for backscattering arising from scalar (i.e. spin-independent) impurities. The  intrinsic spin-orbit coupling and other time-reversal symmetry breaking and/or sublattice-parity breaking potentials also lead  to (sub-leading) corrections to the channel conductance. 
\end{abstract}

\pacs{}
\maketitle

\section{Introduction}
\label{sec:I}
\begin{figure}[b]
\begin{center}
\includegraphics[width=8cm]{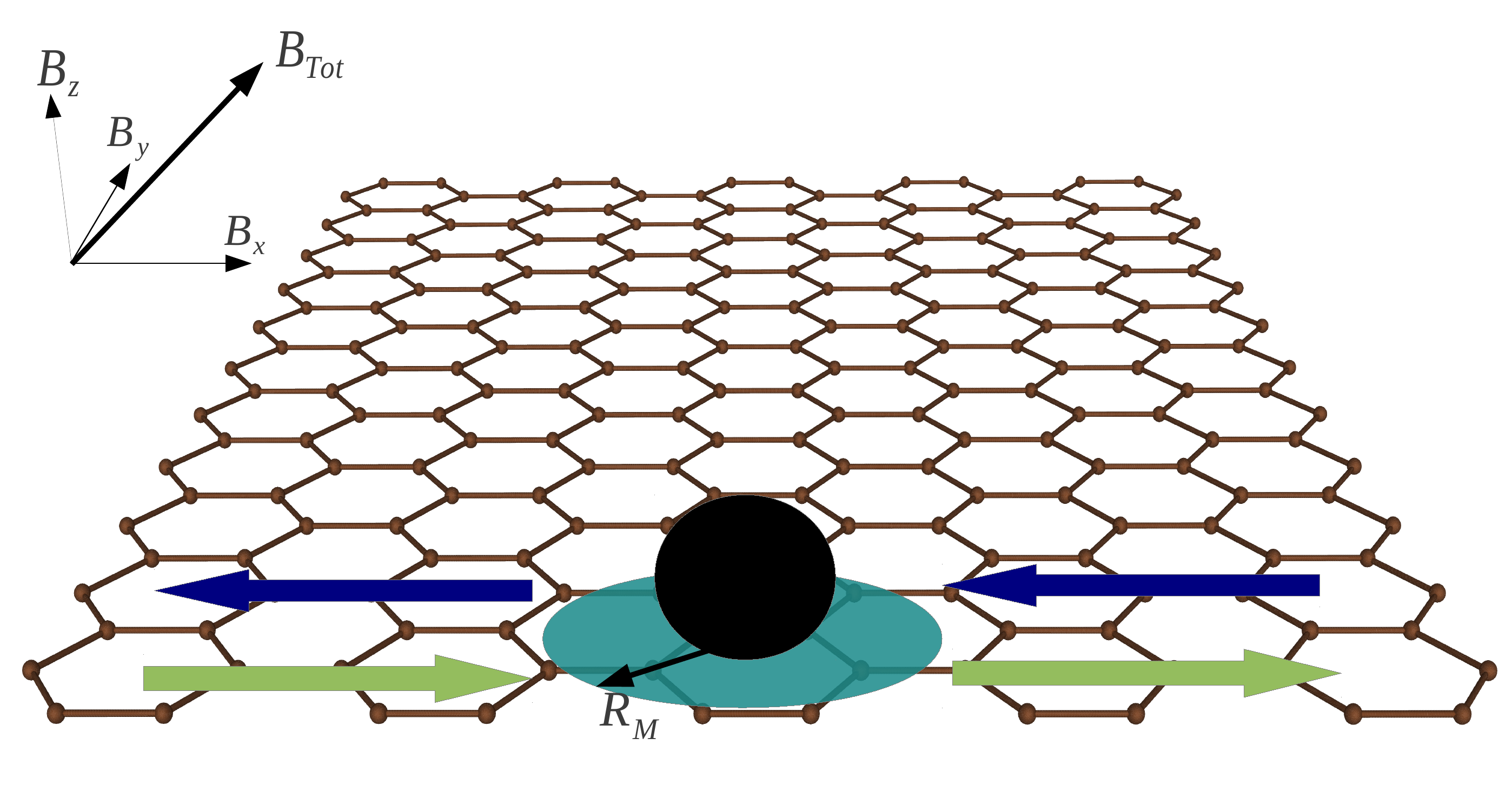}
\end{center}
 \caption{Schematic representation of the system considered in this work: A large metallic cluster (black dot) located near an armchair 
graphene edge induces potential and  spin-orbit coupling by proximity. The size of the cluster is taken to be comparable to the magnetic length ($R_M \approx l_{B}$, not drawn to scale). The cluster induces backscattering between the two counter propagating helical edge channels.}
\label{fig:schematic}
\end{figure}
The quantum spin hall effect (QSHE) \cite{Kane2005,Maciejko2011,KM_Z2top_order,Bernevig15122006,qi2011topological,Konig_QSHE,Kane_Hasan} is an interesting phenomenon related to the emergence of an insulating electronic state in the bulk of a two-dimensional material whose one-dimensional edge remains metallic and exhibits quantized conductance at zero temperature. As first discussed by Kane and Mele for graphene,\cite{Kane2005} in the QSHE state, the helical edges can transport a spin current without dissipation as long as time-reversal symmetry is not broken. Unfortunately, the spin-orbit coupling  is very 
small in graphene~\cite{huertas2006spin,yao2007spin,min2006intrinsic,gmitra2009band} ($\sim 10$ $\mu$eV) 
and therefore the effect is not experimentally accessible in this material.

Nevertheless, in a recent experiment,~\cite{Young2014} transport through helical edges has been observed in monolayer  graphene when the latter is submitted to a strong, tilted magnetic 
field $\vec{B}=(B_{\parallel},0,B_{\perp})$. Under such conditions,  the bulk becomes a magnetic insulator whilst the edge can remain metallic. By increasing the strength of the component  of  $\vec{B}$  parallel to the graphene plane ($B_{\parallel}$), the magnetic state  can be tuned from  anti-ferromagnet (AF), to canted anti-ferromagnet (CAF), and finally to ferromagnet (FM).~\cite{Kharitonov2012,Kharitonov2012a,Young2014} As shown by Kharitonov,~\cite{Kharitonov2012,Kharitonov2012a} (see also Ref.~\onlinecite{Murthy2014}) the nature of the magnetic insulating  is determined by the competition of  the Zeeman and interaction energies. 
At the edge of the magnetic insulator in the CAF and FM states, electrons propagate in opposite directions with (roughly) opposite spins, that is, a helical edge is formed. \cite{Fertig_n_Brey_PRL,Abanin2006,Kharitonov2012,Murthy2014} Although time-reversal symmetry is broken explicitly by the external magnetic field, the two-terminal  conductance of the the helical edge approaches the quantized conductance at the largest magnetic field: The conductance per edge  is $G=G_{0}-\delta G$, where $G_{0}=e^2/h$ and   $\delta G/G_0 \sim 0.1$ at the lowest accessible temperatures in the FM state.~\cite{Young2014}

The physical origin of deviation $\delta G$ is not entirely clear and it is the main purpose of this work to investigate how disorder can  contribute to it. To this end, we assume that a potential scatterer is located near an otherwise perfect armchair graphene edge (see Fig.~\ref{fig:schematic}). Our model is not intended to be a faithful description of the experimental situation found in Ref.~\onlinecite{Young2014}, but we believe that it serves well to the purpose of investigating the effects of 
disorder-induced dissipation at the helical edges. However, if necessary, the model can be easily generalized to extended disorder potentials, a task that we shall not pursue here. 

 Nevertheless, the situation envisaged here can exist in an actual experimental device, where the scatterer may be a contact~\cite{Young2014} or a metallic cluster  located near the edge of the device, for instance.   If the cluster/contact contains heavy atoms (e.g. gold~\cite{Young2014}), the latter can locally induce by proximity a spin-orbit coupling (SOC) as well as providing a source for potential  scattering.~\cite{PhysRevX.1.021001,Ferreira2014,Jaya2014giant,nat_phys_hector} Therefore,
one of the important goals of this work is to investigate the role of the SOC-induced dissipation at the edge. Indeed, in the case of semiconducting quantum wells, it is known that Rashba-like disorder SOC has been identified as one of the possible causes for backscattering,~\cite{Edge_Dynamics_QSH}. In addition,  it was suggested by the authors
of Ref.~\onlinecite{Young2014} that Rashba-like disorder could be an explanation for the absence of perfect quantization of the edge conductance.

In this article, we derive the low-energy effective description of the above model accounting for electron-electron interactions in the helical edge channels, which are crucial in determining the temperature dependence of $\delta G$. The effective low-energy model turns out to be a version of the Kane-Fisher model for an impurity in a Tomonaga-Luttinger
liquid.~\cite{Kane1992} Our analysis shows that the leading contributions to $\delta G$ are potential
scattering and, to a lesser extent, 
the so-called intrinsic (or Kane-Mele) type SOC. However,
we find that, to leading order, the Rashba SOC does not lead to backscattering.  In addition, our analysis indicates that other sources of backscattering sources are potentials that explicitly violate the sublattice inter-change (parity) symmetry and/or time-reversal symmetry. 

  The rest of this article is organized as follows. In 
  Sec. \ref{secII}, we review Kharitonov's results~\cite{Kharitonov2012} for the the energy dispersion of the  armchair edge. Hence, we derive the effective low energy model for the edge states, which  turns out to be, accounting for electron-electron interaction, a  Tomonaga-Luttinger liquid with two counter-propagating edge modes of different Fermi velocity. In Sec. \ref{secIII},  we discuss the properties of a general scatterer, which may correspond to a large metal cluster located near the edge of the sample. In Sec. \ref{secIV}, we obtain the finite temperature conductance  corrections  within linear response theory. A summary and a brief discussion of the possible extensions to this work are provided in Sec.~\ref{secVI}. 
The most technical details of the calculation have been relegated to  the Appendices. In what follows, we work in units where $\hbar=k_{B}=1$.
\section{Low-energy description of the edge}\label{secII}
In order to make the article more self-contained, we will review the main results of Kharitonov's calculation for a semi-infinite graphene 
layer submitted to a strong, tilted magnetic field. Within the $\vec{k}\cdot \vec{p}$ continuum description, the low-energy effective Hamiltonian of graphene under a tilted magnetic field can be written as a sum of three terms, $H=K+V+H_{z}$, where
\begin{align}
K &= \int d\vec{r} \, \Psi^\dagger(\vec{r})  (v_{F} \tau_{z}\sigma_{x}\Pi_{x}+ v_{F} \sigma_{y}\Pi_{y}) \Psi(\vec{r}),
\label{eqn:Kinetic}\\
V &=  \frac{1}{2} \int d \vec{r} d \vec{r}^{\prime} \: 
\Psi^{\dagger}(\vec{r}) 
\Psi^{\dagger}(\vec{r}^{\prime})
 \frac{e^{2}}{|\vec{r}-\vec{r}^{\prime}|} \Psi(\vec{r}^{\prime}) \Psi(\vec{r}),
 \label{eqn:Coulomb}\\
H_{z}&= -\frac{1}{2} g_{s}\mu_{B} \int d\vec{r}\,  \Psi^\dagger(\vec{r}) (\vec{s}\cdot\vec{B}) \Psi(\vec{r});
\end{align}

$K$ is the kinetic energy, $V$ the Coulomb interaction, and $H_{z}$ the Zeeman term; the couplings $g_{s}$, $\mu_{B}$ and $B$ are the Land\'e g-factor, the Bohr magneton and the magnitude of the applied (total) magnetic field, respectively. The Fermi velocity of electrons in graphene is $v_{F}\approx c/300$ ($c$ being the speed of light), and   $\vec{\Pi}=\vec{p} - e \vec{A}(\vec{r})/c$ is the kinetic momentum ($e<0$ is the electron charge).   The matrices $\sigma_{\alpha}$ 
($\tau_{\alpha}$) act upon the sublattice (valley) pseudo-spin and $s_{\alpha}$ act upon the spin degree of freedom. Note that the $z$-axis is perpendicular to the graphene plane and the $x$-axis will be taken along the armchair edge).   The Zeeman energy is proportional to the total magnetic
field $\vec{B} = (B_{\parallel},0,B_{\perp})$, where $B_{\perp}$ ($B_{\parallel}$) is the magnetic field along the perpendicular (parallel) direction to graphene plane.

\begin{figure}[t]
\includegraphics[width=7cm]{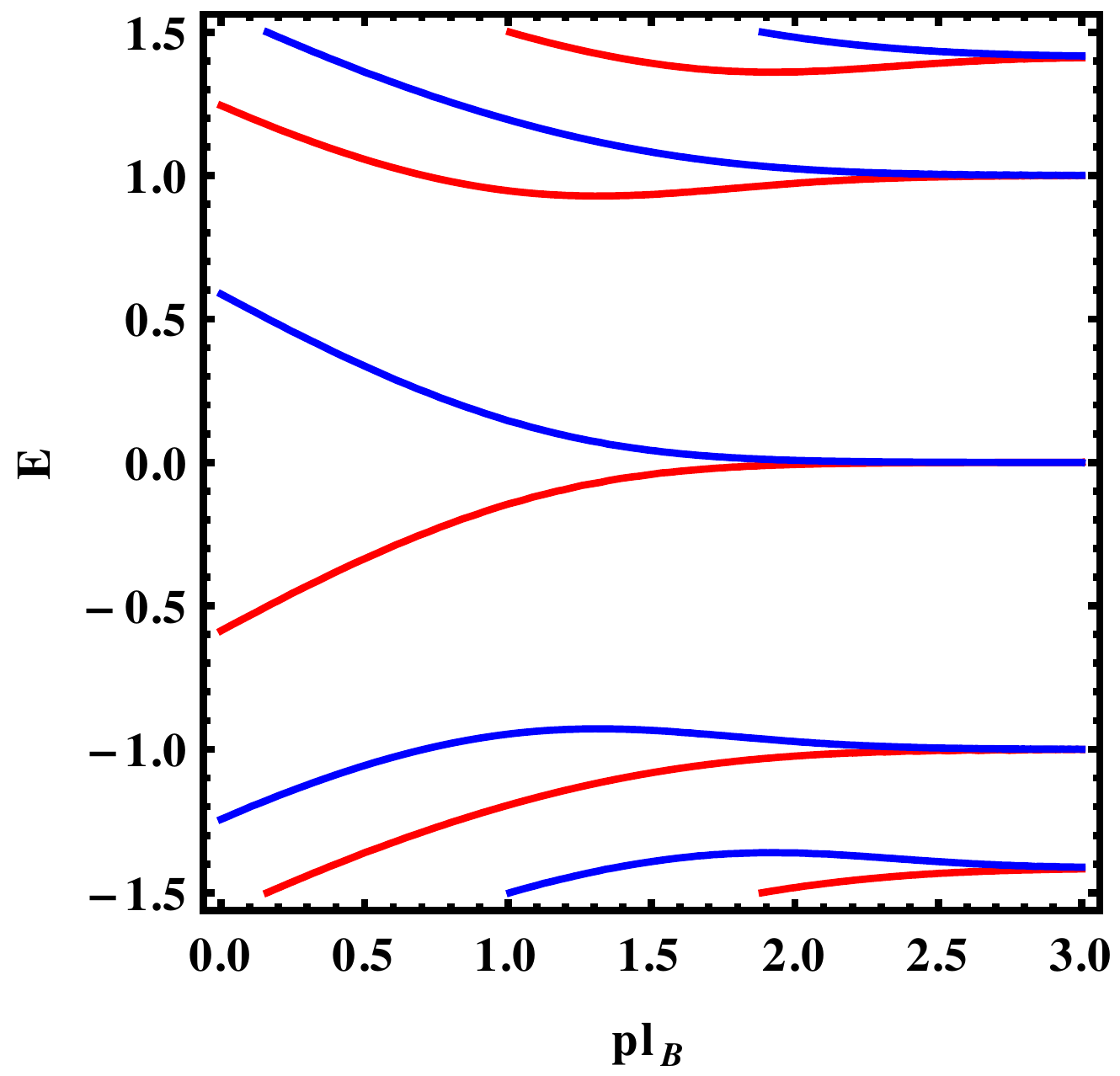}
	\caption{Low energy band structure for non-interacting  electrons in graphene terminated at an armchair edge as af function of $p l_B$ where $p$ is the momentum along the edge (the edge is located at $pl_B \approx 0$) and $l_B$ the magnetic length. Electrons  occupy the region $pl_B \gtrsim 0$. The energy  is in $\sqrt{2}\: v_{F}/l_{B}$ units. The zero Landau level ($E=0$ for $p l_B\to +\infty$) consists of one hole and one electron band, which are split near the edge.}	\label{fig:spectrum_NI}
\end{figure}

Since the largest energy scale of quantum hall system is set by the cyclotron energy, $\omega_{c}= v_{F}/l_{B}$ where $l_{B}$ is the magnetic length, it is useful to neglect the Zeeman and interaction terms in the Hamiltonian and diagonalize the kinetic term first. Furthermore,  in the study of edge effects, two kinds of edge terminations are commonly used, namely the armchair and zig-zag 
edge.~\cite{CastroNeto2009,Carbonin2D,brey2006electronic} In this work, for the sake of simplicity, we use 
armchair edge, which we assume to be located at $y=0$. The energy spectrum of the kinetic energy part of the Hamiltonian for a semi-infinite armchair edge is shown in Fig.~\ref{fig:spectrum_NI}. Note that, within this approximation, all the energy levels spin-degenerate and therefore the ground state is non-magnetic.

  For what follows, it is useful to express the fermionic field operators  $\Psi(\vec{r})$  and $\Psi^\dagger(\vec{r})$ in the basis of Landau level orbitals that diagonalize the kinetic energy $K$:
\begin{equation}
\Psi(\vec{r})=\sum_{n p \tau s}\phi_{n p s}^{\tau}(\vec{r}) c_{n p \tau s}.
\end{equation}
where  $p$ is the momentum along the  edge (i.e. $x$-axis) and $n = 0,  
\pm 1,  \pm 2, \ldots$ is the Landau level quantum number. In the Landau gauge  where $\vec{A}=(B_{\perp}y,0,0)$, 
\begin{align}
\phi_{n p s}^{K}(\vec{r}) &=\frac{e^{i\vec{K}\cdot \vec{r}}}{\sqrt{2}} 
	\begin{pmatrix}
	- \mathrm{sgn}(n) \langle \vec{r} |\, |n|,p\rangle \\
	\langle \vec{r} |\,|n|-1,p\rangle
	\end{pmatrix}
 \otimes \eta_{s},\label{eqn:wavefunction1}\\
\phi_{n p s}^{K^{\prime}}(\vec{r})   &=\frac{e^{i\vec{K^{\prime}}\cdot \vec{r}}}{\sqrt{2}} 
	\begin{pmatrix}
	\langle \vec{r} ||n|-1,p\rangle \\
	\mathrm{sgn}(n) \langle \vec{r} |\, n,p\rangle
	\end{pmatrix}
 \otimes \eta_{s},
\label{eqn:wavefunction2}
\end{align}
for $n \neq 0$ and 
\begin{align}
\phi_{0 p s}^{K}(\vec{r})&=  \frac{e^{i\vec{K}\cdot \vec{r}}}{\sqrt{2}}  \langle \vec{r} |\, n = 0,p\rangle  
	\begin{pmatrix}
	-1\\
	0
	\end{pmatrix}
 \otimes \eta_s,
\label{eqn:wavefunction02}\\
\phi_{0 p s}^{K^{\prime}}(\vec{r}) &=\frac{e^{i\vec{K^{\prime}}\cdot \vec{r}}}{\sqrt{2}}  \langle \vec{r} |\, n = 0,p\rangle
	\begin{pmatrix}
	0\\
	1
	\end{pmatrix}
 \otimes \eta_s,
\label{eqn:wavefunction01}
\end{align}
for $n = 0$. The states $|n, p\rangle$ are the solutions of the Landau level problem of the Schr\"odinger equation in the Landau gauge. In particular, 
\begin{equation}
 \langle \vec{r} |n=0,p\rangle = \frac{e^{ipx}}{\sqrt{L_{x}}} 
 \frac{\exp\left( - \frac{(y-p l_{B}^2)^2}{ 2l_{B}^{2} }  \right)} { \left(\pi l_{B}^2\right)^{1/4} }.
\end{equation}
where $l_B = \sqrt{\frac{c}{eB_{\perp}}}$ is the magnetic length and $L_x$ the length of the (armchair) edge. Finally, the spinors:
\begin{equation}
\eta_{\uparrow} = \left( \begin{array}{c}  1 \\ 0\end{array} \right) \quad  
\eta_{\downarrow} = \left( \begin{array}{c}  0 \\ 1\end{array} \right).
\end{equation}

 Note that in the $n = 0$ Landau level (0LL), the sublattice and valley (pseudo-)spins are aligned. This feature will probe extremely important in the  discussion that follows. Accounting for the spin degeneracy, each single-particle state in the 0LL is four-fold degenerate corresponding to $s = \uparrow,\downarrow$
and $\tau = K, K^{\prime}$.

\subsection{Zeeman and interaction energy effects}
In order to fully determine the ground state,  we next take into account the Coulomb interaction and the Zeeman term. In  neutral (i.e. undoped) graphene, the Landau level of zero energy (0LL) is half filled and therefore all Landau levels below (above) 0LL are completely filled (empty). At temperatures far below the cyclotron energy (i.e. $T\ll \omega_c$) and close to half-filling,  the low energy dynamics is determined by the electrons in the 0LL. As a result, we project the Coulomb interaction (Eq.  \ref{eqn:Coulomb}) onto the subspace of states belonging to the 0LL. Thus, in what follows, we shall suppress the Landau level
index $n$ as it is implicitly understood that $n  =0$. In addition, we will suppress 
the sublattice index because in the 0LL the pseudo-spin and valley spin are aligned.
The Coulomb interaction in the 0LL subspace is therefore formally described by~\cite{Kharitonov2012} 
\begin{align}
V_{0LL}&=\frac{1}{2}\sum_{\alpha=0,x,y,z} \sum_{p_{1}...p_{4}} u_{\alpha}(p_1,p_2;p_3,p_4) \: :T^{\alpha}_{p_2, p_1}  T^{\alpha}_{p_3,p_4}:  \notag \, ,\\
T^{\alpha}_{p_1,p_2} &=  \sum_{s}\sum_{\sigma,\sigma^{\prime}}  c_{p_{1}\sigma s}^\dagger (\tau_{\alpha})_{\sigma \sigma^{\prime}}c_{p_{2} \sigma^{\prime} s}\, ,
\label{eqn:Effective Coulomb}
\end{align}
where $:...:$ stands for normal ordering. The Coulomb amplitudes $u_{\alpha}$  have been
discussed in Refs.~\onlinecite{Goerbig2011,Kharitonov2012} and we will not dwell on them here. Nevertheless,  it is worth mentioning that the Coulomb amplitudes that do not conserve valley quantum number are exponentially suppressed by an exponential factor $\sim e^{-(l_{B}/a)^2}$ where $a$ is the distance between the carbon atoms in graphene. Thus, the intra-valley amplitude,  $u_{0}$, is dominant one.~\cite{Goerbig2011} In fact, if only $u_{0}$ is retained in Eq.~\eqref{eqn:Effective Coulomb}, the Hamiltonian  and the ground state would exhibit SU$(4)$ symmetry (see Ref.\onlinecite{Goerbig2011} and references therein).

  However, when the sub-leading amplitudes  $u_{\alpha=1,2,3}$ that violate the SU$(4)$ symmetry are also included, together with the Zeeman energy, a quantum hall magnet can be stabilized as the ground state.~\cite{QHF_bilayer_system,QHF_graphene,Kharitonov2012}  Its magnetic order can be uncovered by relying upon mean-field theory for which the order parameter is $P_{\tau s, \tau^{\prime} s^{\prime}} = \langle 
c^{\dag}_{p\tau s} c_{p \tau^{\prime} s^{\prime}} \rangle$.~\cite{Kharitonov2012,Kharitonov2012a} 
This order parameter can be  written  as follows:
\begin{equation}
P=\chi_{a}\chi_{a}^{\dagger}+\chi_{b}\chi_{b}^{\dagger},
\end{equation}
where $\chi_{a}=|K\rangle \otimes |S_{a}\rangle$ and $\chi_{b}=|K'\rangle \otimes |S_{b}\rangle$ are the trial spin-valley spinors. For the semi-infinite system, we shall neglect any deviation of the order parameter $P$ from its bulk value near the edge.~\cite{Fertig_n_Brey_PRL,PhysRevB.79.115431,Kharitonov2012}  
We shall also neglect the 
dynamics of the order parameter, which
can lead to additional  dissipation mechanisms in clean systems (via coupling with the bulk Goldstone modes), different from those discussed here.~\cite{Murthy2014}
In spin space, the spinor accepts the following
parametrization:
\begin{equation}
|S_{a,b}\rangle=\begin{pmatrix}\cos\ \frac{\theta_{s}}{2}\\
\pm \sin\frac{\theta_{s}}{2} e^{i\phi_{s}}
\end{pmatrix}
\end{equation}
where $\theta_{s}$ is the angle between the spin polarization and the total magnetic field $\vec{B}$; $\phi_{s}$ is the azimuthal angle around the total magnetic field. The positive sign applies to $S_a$ and the negative sign to $S_b$. \cite{Kharitonov2012a,Kharitonov2012} 
The above parametrization assumes that the spin-quantization axis points along the direction of the total magnetic field $\vec{B}$. Note that this choice is different from the one assumed for the operator $s_{z}$, which measures the spin-projection perpendicular to the plane of graphene and $s_{x},s_{y}$, which measure the in-plane spin projection. Both sets of spin operators $s_{\alpha}$
and $s^{\prime}_{\alpha}$ ($\alpha = x, y, z$) are related by a rotation:
\begin{align}
s^{\prime}_{z} &= s_{z}\cos\theta + s_{x}\sin\theta, \label{eq:spz}\\
s^{\prime}_{x} &= -s_{z}\sin\theta + s_{x}\cos\theta,\\
s^{\prime}_{y} &= s_{y}.
\end{align}
\begin{figure*}[t]
\includegraphics[scale=0.5]{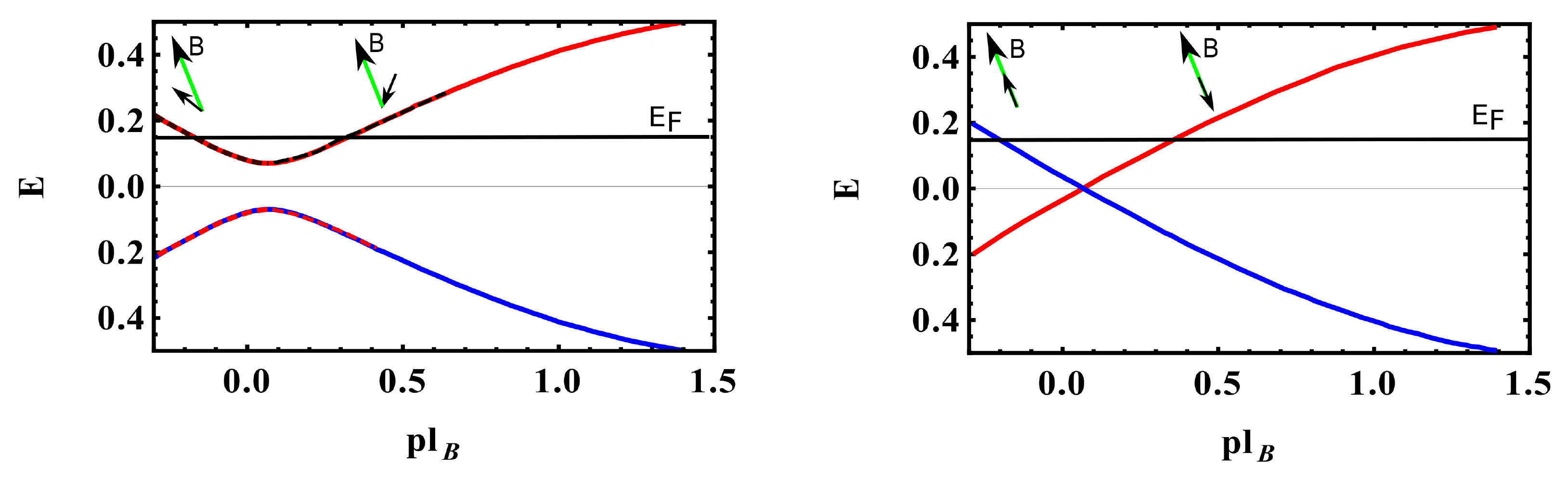}
\caption{Lowest energy band structure of a graphene armchair edge within mean field theory for the canted anti-ferromagnetic (CAF, left panel) and ferromagnetic (FM, right panel), respectively. The edge is located at $pl_{B} \approx 0$. As a result of the Zeeman and the Coulomb interaction, an energy gap develops in the band structure, which is minimum near the edge. In electron doped graphene, the low energy dynamics are governed by the degrees of freedom near the Fermi points where the upper band intersects with the Fermi energy $E_F$ (for hole doping $E_F$ is within the lower band).}
\label{fig:spectrum_I}
\end{figure*}

 By minimizing the kinetic energy and Zeeman term together with the Coulomb interaction within the mean-field approach,  the single-particle spectrum shown in Fig. \ref{fig:spectrum_I} is obtained.~\cite{Kharitonov2012}  The Zeeman term in the Hamiltonian lifts the degeneracy between the spin up and spin down bands. However, due to the exchange part of the Coulomb interaction, the 0LL bands near the edge anti-cross (for $pl_B\approx0.25$) and develop an energy gap. The magnitude of the gap is given by,~\cite{Kharitonov2012}
\begin{equation}
\Delta_{\mathrm{edge}}=\begin{cases}
\Delta \sqrt{1-(\frac{\epsilon_{z}}{2|u_{1}|})^{2}} & \theta<\theta_{\mathrm{cr}}\quad \mathrm{(CAF)}\\
0 &  \theta\geq\theta_{\mathrm{cr}}\quad \mathrm{(F)}
\end{cases}
\end{equation}
where $\theta=\tan^{-1}(B_{\parallel}/B_{\perp})$ is angle subtended by $\vec{B}$ with the $z$-axis perpendicular to the graphene plane, $\epsilon_{z}=\frac{1}{2}g_{s} \mu_{B} B $ is the Zeeman energy, and $\Delta=u_{0}+2|u_{1}|+u_{3}$; $\theta_{\mathrm{cr}}=\cos^{-1}\left(\epsilon_{z}/2|u_{\perp}|\right)$ is the critical angle for which a quantum phase transition from the canted anti-ferromagnet (CAF) and ferromagnet (FM)  takes place when the Zeeman energy takes over the exchange energy. 

In the CAF state, when the chemical potential lies above (below) the gap in the upper (lower) band,  two metallic channels appear at the edge (corresponding to the two Fermi points in Fig.~\ref{fig:spectrum_I} where the energy band intersects the chemical potential). The corresponding states are not eigenstates of $s^{\prime}_{z} = \vec{s}\cdot \vec{B}/B$ and are not mutually orthogonal. This has a significant effect on the suppression of conductance from its perfect quantized value $G_0$, as we will  see in Sec. \ref{secIII}. 

 As the angle $\theta$ is tuned across its critical value, $\theta_{\mathrm{cr}}$, and the system becomes a FM, the two conducting channels along the edge still exists. Unlike in the CAF, they are eigenstates of $s^{\prime}_{z}$ (cf. Eq.~\ref{eq:spz}) and therefore mutually orthogonal. Nevertheless, we note that this is the result of the approximation made in Ref.~\onlinecite{Kharitonov2012}, which neglects the variation of the order parameter near the edge. The more sophisticated
treatment provided in Ref.~\onlinecite{Murthy2014} indicates that
spin orientation of the edge states remains slight canted in the FM, only becoming perfectly anti-aligned at large values of $B_{\parallel}$. 
However,  to the extent they can be assumed to be perfectly spin anti-aligned, the FM edge channels can transport spin and charge current without dissipation if there were no impurities. However, different  from the QSHE in semiconducting quantum wells, the QSHE in the FM  quantum hall insulator is not protected by time-reversal symmetry but the U$(1)$ symmetry generated by the total $s^{\prime}_{z}$.
 
 In both the CAF and FM states, at  $T\ll \min\{\Delta_{\mathrm{edge}}, E_F\}$, the low-energy 
 physics is controlled by the fermionic degrees of freedom near the Fermi energy $E_F$. Upon linearlizing the band around the Fermi points defined by the intersects 
($p_{F,L}$ and $p_{F,R}$)  of the bands with $E_F$ (see Fig. \ref{fig:spectrum_I}), an effective low-energy Hamiltonian for the edge modes can be obtained:
\begin{equation}
H_{0} =\sum_{c=R,L}\sum_{p} v_{c}\: p\,  :\psi_{c}^{\dagger}(p)\psi_{c}(p):\, ,
\end{equation}
where $\psi_{R}(p) \sim  c_{p_{F,R}+p}$ and  $\psi_{L}(p) =  c_{p_{F,L}-p}$ for $|p| \ll \Lambda$ and $\Lambda \sim E_F/\max\{v_{R},v_{L}\}$ and
\begin{equation}
v_{R}=\frac{dE(p)}{dp}\Big|_{p = p_{F,R}} \qquad
v_{L}= -\frac{dE(p)}{dp}\Big|_{p = p_{F,L}}.
\end{equation}
The creation and annihilation operator obey the usual fermion anti-commutation relation $\{\psi_{c}(p),\psi_{c'}^{\dagger}(p')\}=\delta_{p,p'}\delta_{c,c'}$ 
where $c=(R,L)$.  Note that, unlike the QSHE in semiconductor quantum wells for which the states at the two channels are a Kramer's pair,~\cite{Edge_Dynamics_QSH} in the present case the right and left movers can have different Fermi  velocities  (i.e. $v_{R}\neq v_{L}$). 

 In addition, since the degrees of freedom near the edge are gapless, we need to take into account the
effect of electron-electron interactions.  In experimental conditions, due to the presence of a nearby metallic gate used to tune the Fermi energy $E_F$, the Coulomb interaction between electrons is screened to yield a short-range interaction:
\begin{equation}
V_{\mathrm{edge}}=2\pi g_{RL}\int dx\,  :\psi_R^\dagger(x)\psi_R(x)\psi_L^\dagger(x)\psi_L(x):\, .
\end{equation}
Note that $V_{\mathrm{edge}}$ is not necessarily repulsive
as the Coulomb interaction may be over screened at low energies and become an effective attractive
interaction. As a consequence, below we have considered both signs for $g_{RL}$. Thus, the complete low-energy Hamiltonian for a clean armchair edge reads:
\begin{equation}
H_{\mathrm{edge}} =  H_0 + V_{\mathrm{edge}}.
 \label{eq:H_{edge}}
\end{equation}

\section{Scattering Potentials} \label{secIII}

\subsection{General form of a localized scatterer potential}
We next analyse how the presence of a scatterer affects conduction along the edge.  To this end, it is instructive to consider the most general form of the potential created by a generic scatterer located near the edge of the sample. Generally, the latter can be written as follows: 
\begin{equation}
V_{M}= \int d\vec{r} \, \Psi^\dagger(\vec{r}) V(\vec{r})\Psi(\vec{r}),
\end{equation}
where $V(\vec{r})$ is a $8 \times 8$ potential matrix acting on both sublattice, valley, and spin degrees of freedom. Since a tilted magnetic field  break most symmetries of the graphene Hamiltonian, the most general form of $V(\vec{r})$ is a linear combination of the $64$ matrices $\tau_{\alpha} \sigma_{\beta}s_{\gamma}$, where $\alpha,\beta,\gamma = 0,x,y,z$ ($0$ referring to the identity matrix):
\begin{equation}
V(\vec{r}) = \sum_{\alpha,\beta,\gamma} V_{\alpha\beta\gamma}(\vec{r})\tau_{\alpha} \sigma_{\beta}s_{\gamma}, \label{eq:genimp}
\end{equation}
where the relative strength of each potential 
 $V_{\alpha\beta\gamma}(\vec{r})$ depends on the microscopic details of the scatterer, the strength of the external magnetic field $B_{\perp}$ and $B_{\parallel}$, as well as other perturbations (boundary conditions at the edge). Therefore, $V_M$ will contain
 terms that break the symmetries of the zero-field Hamiltonian. 
Those terms are generated by the projection onto the low-energy subspace
containing the zero energy Landau levels.   In what follows, we shall discuss which terms are expected to yield a dominant contribution to $V_M$. Rather than trying to be exhaustive, our choice will be physically motivated. However, we shall first take a detour to discuss some general properties of the 0LL projection.

\subsection{Projection of sublattice-valley Operators onto the 0LL}
\label{sec:project}

  In order to obtain a low-energy effective description of the edge channels in the presence of the scatterer, its potential must be projected onto the 0LL. To leading order, the scatterer potential in the 0LL is given by
\begin{equation}
V_{0LL} =  \mathcal{P}V(\vec{r}) \mathcal{P} + \cdots 
\end{equation}
where $\mathcal{P}$ is the projection operator in the 0LL. The terms contained in the ellipsis represent corrections arising from virtual transitions to Landau levels with $n\neq 0$.  Since the scatterer potential introduced above acts upon the sub-lattice, valley and spin degrees of freedom, it is worth dwelling a bit on the explicit form of the 0LL projection operator $\mathcal{P}$   and how it affects the Pauli matrices $\tau_{\alpha}\sigma_{\beta}s_{\gamma}$. 
  
   The operator $\mathcal{P}$  can can be written as follows:
\begin{align}
\mathcal{P} &= \mathcal{P}_{sv}\otimes  \mathcal{P}_o, \label{eq:project}\\
P_{sv} &= |A K\rangle \langle A K | + |B K^{\prime} \rangle \langle B K ^{\prime} | ,\\
\mathcal{P}_o  &= |n = 0, p\rangle \langle n= 0, p|
\end{align}
The first term in the right hand-side of  Eq.~\eqref{eq:project} ($\mathcal{P}_{sv}$) acts on the sub-lattice
and valley pseudo-spin degrees of freedom, whereas the second term ($\mathcal{P}_o$) acts on the other orbital degrees of freedom. Since sub-lattice spin and valley spin are aligned in 0LL,  when projecting the Pauli matrices using $P_{sv}$, we obtain the following:
\begin{align}
\mathcal{P}_{sv}\tau^0\sigma_{z}\mathcal{P}_{sv} &= \Sigma_z,\quad 
\mathcal{P}_{sv}\tau_z\sigma^{0}\mathcal{P}_{sv} = \Sigma_z,\\
\mathcal{P}_{sv}\tau^{-}\sigma^{+}\mathcal{P}_{sv} &= \Sigma^{-},\quad
\mathcal{P}_{sv}\tau^{+}\sigma^{-}\mathcal{P}_{sv} = \Sigma^{+},\\
\mathcal{P}_{sv}\tau_z\sigma_{z}\mathcal{P}_{sv} &= \mathcal{P}_{sv},\quad 
\mathcal{P}_{sv}\tau^{\pm}\sigma_{0}\mathcal{P}_{sv} = 0,\\
\mathcal{P}_{sv}\tau_0\sigma^{\pm}\mathcal{P}_{sv} &= 0,\quad
\mathcal{P}_{sv} \tau_z\sigma^{\pm}\mathcal{P}_{sv} = 0, \\
\mathcal{P}_{sv} \tau^{\pm} \sigma_z \mathcal{P}_{sv} &= 0,\quad
\mathcal{P}_{sv} \tau^{+}\sigma^{+} \mathcal{P}_{sv} = 0,\\
\mathcal{P}_{sv}\tau^{-}\sigma^{-}\mathcal{P}_{sv} &= 0.
\end{align}
where we have introduced the operators:
\begin{align}
\Sigma_z &= |A K\rangle \langle A K | - |B K^{\prime} \rangle \langle B K ^{\prime} |,\\
\Sigma^{+} &= |B K\rangle \langle A K^{\prime} |, \quad
\Sigma^{-} = |A K^{\prime}\rangle \langle B K |.
\end{align}
Thus, $\{\mathcal{P}_{sv},\Sigma_z,\Sigma^{+},\Sigma^{-}\}$ behave  effectively as Pauli matrices in the subspace of 0LL states.

\subsection{Time-Reversal Invariant Scattering Potentials}\label{sec:trs}
Although the most general impurity potential should not respect time-reversal symmetry  as the latter is broken by the external magnetic field,
it is worth beginning our analysis by considering  various time-reversal symmetric (TRS) potentials. The reason is that TRS  potentials are present in the microscopic Hamiltonian of graphene with impurities at zero field.~
 However, non-TRS potentials need to be generated by means of virtual transitions to higher energy states in the presence of the time-reversal symmetry breaking perturbations like the magnetic field. Therefore, according to the discussion in previous section, TRS potentials are expected to have  larger strength relative to the non-TRS ones. In addition, the discussion that follows is also motivated by the suggestion made in Ref.~\onlinecite{Young2014} attributing the absence of conductance quantization to the existence of scatterers proximity-induced Rashba
spin-orbit coupling (SOC).~\cite{PhysRevX.1.021001,Ferreira2014}  

  TRS potentials can be divided into two classes depending on  how they act on the spin degree of freedom: Those proportional to $s_0$, which are termed `scalar' potentials, and those proportional to $s_{x,y,z}$, which correspond to SOC. These two classes can be further subdivided into those inducing only intra-valley scattering and those inducing intervalley scattering depending on whether they commute with $\tau_z$ or not. Note that, typically, the strength
of the scalar potentials is much larger than the strength of SOC potentials~\cite{PhysRevX.1.021001,Ferreira2014, Jaya2014giant}. 
  
  An example for a TRS scalar potential is:
\begin{align}
V_{\mathrm{s}}(\vec{r}) = V_{0}(\vec{r})s_0  + V_{x}(\vec{r}) \tau_x  s_0 \notag \\
 + V_{xx}(\vec{r}) \sigma_x \tau_x s_0 +\cdots  
\label{eq:g_scalar}
\end{align}
where the second and third terms $[\propto V_{x}(\vec{r}), V_{xx}(\vec{r})]$  and others similar to them not explicitly written describe intervalley scattering. Strong inter-valley scattering requires that a rapid spatial variation of  $V_{x}(\vec{r})$ and $V_{xx}(\vec{r})$ on the scale of the inter-carbon distance $a$ in graphene.
In other words, it requires atomic scale disorder (i.e. point defects, edge roughness, etcetera). It is possible to estimate, within the Born approximation, the reflection coefficient at the edge from such
small-size scatterers,~\cite{GKG2010} which is $\sim (e^2/v_F) (a/l_B)^2$. However, for the  experimentally accessible values of $B_{\perp}$,  $l_B \approx 26/\sqrt{B_{\perp}(T)}$ nm $\gg a=0.24$ nm, implying that the intervalley scattering terms in the above equation can be neglected. For the same reason, we shall also neglect the inter-valley SOC 
scattering terms below. 

 Next, we consider the (intra-valley) SOC scattering potentials, which can take the following general form:
\begin{align}
V_{\mathrm{so}}(\mathrm{r}) &= V_{\mathrm{iso}}(\vec{r}) \sigma_z \tau_z 
s_z + V_{\mathrm{pbiso}}(\vec{r}) \tau_z s_z \notag \\
&+ V_{\mathrm{R}x}(\vec{r}) \sigma_x\tau_z s_y  - V_{\mathrm{R}y}(\vec{r})  \sigma_y s_x,
\label{eq:g_soc}
\end{align}
where the first term on the right hand-side corresponds to the  intrinsic (or Kane-Mele) type of SOC, and
the third and fourth terms correspond to the Rashba-like SOC. Note that we have considered a generalized anisotropic Rashba, for which $V_{\mathrm{R}x}(\vec{r})\neq V_{\mathrm{R}y}(\vec{r})$. 
Such SOC potentials can be induced when the 
hexagonal point-group symmetry of graphene is broken down to a smaller symmetry group by proximity to substrates and clusters of 
heavy-metals.~\cite{nat_phys_hector,Ochoa2012} The second term
$\propto V_{\mathrm{pbiso}}(\vec{r})$ is absent in 
the microscopic Hamiltonian because it is forbidden by the parity symmetry that exchanges of the A and B sub-lattices. However, it can be generated by the application  of parity-breaking fields like a tilted magnetic field or by the edge potential. Since this term is not present at the zero-field (for which parity symmetry is not broken), in a first approximation, we  shall neglect it compared to the scalar and standard SOC terms. However, in the following section we shall see that this term can yield backscattering in the FM state.

\begin{table}[t]
\begin{ruledtabular}
\begin{tabular}{cc}
Scattering Operator  & 0LL Projection \\
\hline\\
$\tau_{0}\sigma_{0} s_{0}$ &  $\mathcal{P}_{sv}\: s_{0}$ \\
$\sigma_{0}\tau_{x} s_{0}$ & 0\\
$\sigma_{x}\tau_{x} s_{0}$ & $\Sigma_x \: s_0$\\
$\sigma_{z} \tau_{z}s_{z}$  & $ \mathcal{P}_{sv}\:  s_{z}$ \\
$\sigma_{0}\tau_z s_z$ & $\Sigma_z s_z$ \\
$\sigma_{x}\tau_{z}s_{y}$ & $0$  \\
$\sigma_{y}\tau_{0}s_{x}$ & $0$  \\
\end{tabular}
\end{ruledtabular}
\caption{\label{tab:table1}
Projection of several types of time-reversal symmetirc scattering potentials onto the zero energy Landau Level of graphene (0LL).
We have introduced the operator $\Sigma_x = (\Sigma^{+}+\Sigma^{-})/2$}
\end{table}

 Upon projecting \eqref{eq:g_scalar} and \eqref{eq:g_soc} onto the 0LL, 
we obtain: 
\begin{equation}\label{eq:soc_eff}
\mathcal{P}V^{TRS}_{M}\mathcal{P} = \int d\vec{r} \, 
\Psi^{\dag}_{0}(\vec{r})\left[ V_{0}(\vec{r}) s_0 + V_\text{{iso}}(\vec{r})s_{z} \right] \Psi_{0}(\vec{r}), \qquad
\end{equation}
where   $\Psi_{0}(\vec{r}) = \sum_{p\tau s} \phi^{\tau}_{0p s}(\vec{r}) c_{p\tau s}$ ($\Psi_{0}^\dagger(\vec{r})  = \left[\Psi_0(\vec{r})\right]^{\dag}$) is the annihilation (creation) field operator in the 0LL, respectively.  Note that, in the above expression, $s_{z}$  measures the projection of the electron spin along the axis perpendicular to the  graphene layer and must be expressed in terms of $s^{\prime}_x$ and $s^{\prime}_{z}$  (cf. Eq.~\eqref{eq:spz}).

It is noteworthy that Rashba SOC of the form given in  Eq.~\eqref{eq:g_soc} gives no contribution to the 0LL projected $V_{M}$ (cf. table~\ref{tab:table1}). Therefore, it alone cannot  suppress the conductance  from its perfect quantized value, unlike the case of the QSHE in two-dimensional TRS topological insulators.~\citep{Edge_Dynamics_QSH,PRL_Moore}  This is a direct consequence of the  structure of 0LL orbitals in graphene, for which the sub-lattice and valley pseudo-spins are locked. Landau level mixing arising from interactions, etc, can modify this conclusion slightly, but the corrections are expected to be small and suppressed at the high magnetic fields. On the other hand, the scalar and intrinsic SOC potential are not affected by the 0LL projection and therefore can lead to backscattering.

\begin{figure}[t]
\includegraphics[width=8cm]{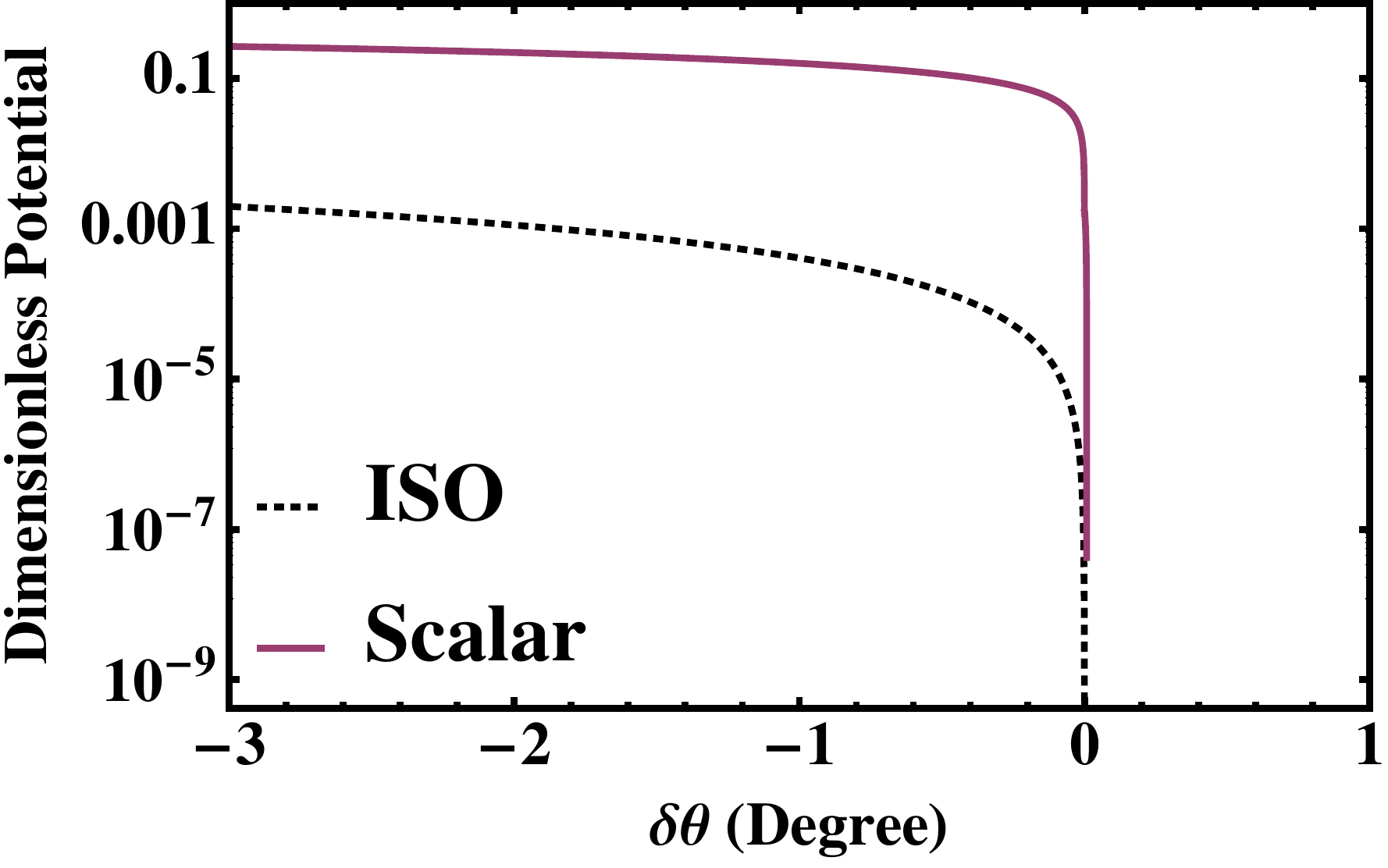}
\caption{Normalized strength  of the leading time-reversal invariant contributions to scatterer potential. The intrinsic SOC (dashed line), $g_{\mathrm{iso}}/V_{\mathrm{iso}}$, and scalar potential (solid line), $g_{s}/V_0$, can be induced by proximity to a heavy metal cluster. Their strength is plotted against the angular deviation $\delta\theta=\theta-\theta_{cr}$ (in degrees);  $\theta_{cr}$ is the critical polar angle of the magnetic field for which quantum phase transition to ferromagnet (FM) occurs. The vanishing of the scattering potential strength in Ferromagnet (FM) is an artefact of the assumption that the
scattering potential is entirely due to a combination of scalar 
and intrinsic (i.e. Kane-Mele type) spin-orbit coupling (SOC) along with
the assumption of that the order parameter does not deviate from its bulk value.  
Note that, in the canted anti-ferromagnet corresponding to $\delta \theta < 0$, the contribution of the scalar potential is larger than the SOC contribution.}
\label{fig:g}
\end{figure}

To make further progress towards obtaining  the effective Hamiltonian of a 
TRS scatterer at the edge,  let us assume the size of the scatterer (e.g. 
a metal cluster of radius $R_{M}$) to be comparable to the magnetic length,
$R_{M}\sim l_B$. Thus, we can approximate  
$V_0(\vec{r})$ and $V_{\mathrm{iso}}(\vec{r})$  
by Dirac-delta functions, which, upon further projection
on the low-energy modes near the edge, leads to
\begin{equation}
V_{M} =  g R_{M} \left[ \psi_{R}^{\dagger}(0) \psi_{L}(0)+ \psi_{L}^{\dagger}(0) \psi_{R}(0)\right].
\label{eqn:H}
\end{equation}
where $g$ has energy units.  When combined with Eq.~\eqref{eq:H_{edge}}, the Hamiltonian, $H = H_{\mathrm{edge}} + V_{M}$, is a version of the model studied by Kane and Fisher~\cite{Kane1992} for an impurity in a Tomonaga-Luttinger liquid. The only difference is that, in our model, right and left movers have different  Fermi velocities. Note that, after projecting $V_M$ on the edge,  a term of the form $\psi_{L}^{\dagger}(x) \psi_{L}(x) + \psi_{R}^{\dagger}(x) \psi_{R}(x)$ has been dropped since it does not contribute to the channel resistance.~\cite{Kane1992,Gia_QPin1D} The scattering potential coupling $g$ contains contributions from both scalar and the intrinsic SOC potentials, i.e. $g=g_{s}+g_{iso}$, where
\begin{align}
g_{s}&=\frac{\sqrt{2\pi}  V_{0}}{\sqrt{N(p_{R})N(p_{L})}} \left[\frac{A(p_{R})A(p_{L})}{\Delta_{AF}^{2}}+1 \right],\\
g_{iso}&=\frac{\sqrt{2\pi}  V_{\mathrm{iso}} \cos\theta}{\sqrt{N(p_{R})N(p_{L})}} \left[\frac{A(p_{R})A(p_{L})}{\Delta_{AF}^{2}}-1 \right] ,
\end{align}
with 
\begin{align}
A(p) &= \Delta_{F}+\epsilon_{Z}-\epsilon(p) \notag \\
& \quad - \sqrt{\Delta_{AF}^{2}+[\epsilon(p)-(\Delta_{F}+\epsilon_{Z})]^{2}}\quad;
\end{align}
$\epsilon(p)$ is the single particle energy, which is displayed in Fig. \ref{fig:spectrum_NI} and $N(p)=\sqrt{A(p)^{2} +\Delta_{AF}^2 }$ is the normalization of the single particle wave function; $\theta= \tan^{-1}\left( B_{\parallel}/B_{\perp} \right)$ is the polar angle of the applied magnetic field, $\Delta_{F}=\frac{1}{2}(u_{0}-2|u_{1}|+u_{3})\cos\theta_{s}$ and $\Delta_{AF}=\frac{1}{2}(u_{0}+2|u_{1}|+u_{3})\sin\theta_{s}$ are the combination of the mean-field parameters that favour ferromagnetic and anti-ferromagnetic order, respectively. 
The relative strength of these potentials $g_s/V_0$ and 
$g_{\mathrm{iso}}/V_{\mathrm{iso}}$ has been plotted 
in Fig.~\ref{fig:g}. Note that $g_s/V_0 \gg g_{\mathrm{iso}}/V_{\mathrm{iso}}$. In addition, 
typically,~\cite{Ferreira2014,Jaya2014giant}
$V_{0}\gg V_{\mathrm{iso}}$, which means that the
main contribution stems from the scalar part of the 
scatterer potential.
\subsection{Backscattering in the Ferromagnet}\label{subsec:FM}
As shown in Fig. \ref{fig:g}, the effective potential strength arising from the
scalar potential and intrinsic SOC vanishes in the FM state.  However,  to understand the vanishing of $g_{s},g_{\mathrm{iso}}$ better, let us first recall the valley and spin structure of the single-particle orbitals in the FM state:
\begin{align}
|R \rangle &=\left(\frac{|A K\rangle+|B K'\rangle}{2} \right) \otimes |\uparrow\rangle,\label{eq:l1}\\
|L \rangle &= \left( \frac{|A K\rangle-|B K'\rangle}{2} \right) \otimes |\downarrow\rangle.\label{eq:r1}
\end{align}
Note that the in the FM the orbitals are eigenvectors of $s^{\prime}_z$ and
therefore orthogonal. This means that no scalar (intra- or inter-valley) is effective in causing backscattering. Thus, only spin-dependent potentials can cause backscattering. However, the orbital part of the intrinsic SOC ($\sim\sigma_z \tau_z s_z$), cannot couple two edge 
states even if the spin operator does (recall that they are eigenstates of $s^{\prime}_z$, not $s_z$).

However, we need to stress that the above conclusions rely on the orbital structure of the single-particle orbitals in the FM described
by Eqs.~\eqref{eq:r1} and \eqref{eq:l1}.  The structure is the result of the assumption (made in Ref.~\onlinecite{Kharitonov2012} and which we follow here) that the magnitude of the  order
parameter does not deviate from its bulk value. This approximation is expected to be accurate~\cite{Murthy2014}
for large values of $B_{\parallel}$. However, 
a more sophisticated mean-field treatment like the one provided in Ref.~\onlinecite{Murthy2014}, which allows for the spatial variation of the order parameter, would not automatically imply a vanishing impurity (back)scattering in the FM. The reason is that, near the edge, the order parameter can deviate from its bulk value and, as a consequence, the spins near the edge of the FM can be slightly canted.~\cite{Murthy2014}  This implies that the backscattering due to the scalar and intrinsic SOC potentials is suppressed relative to its values in the CAF but not they are completely eliminated. On the other hand, a Rashba SOC potential yields no backscattering to leading order. The latter result
is independent of detailed form of the single-particle orbitals as it is entirely a consequence of the 0LL projection.

Nevertheless, it is also interesting to consider additional scattering potentials that have been neglected so far, which can also lead to backscattering.  If we restrict ourselves to TRS potentials, as we did in the previous section, one possible candidate, which by no means exhausts all the possibilities, is the following parity-breaking intrinsic SOC:
\begin{align}
V^{TRS}_{M} &= \int d\vec{r}\:  V_{\mathrm{pbiso}}(\vec{r}) \Psi^{\dag}(\vec{r})\tau_z s_z\Psi(\vec{r}) +\cdots  \\ 
&=  \int d\vec{r}\:  V_{\mathrm{pbiso}}(\vec{r}) \Psi(\vec{r}) \tau_z 
\left( s^{\prime}_z \cos \theta   \right. \notag \\
& \quad \left. -  s^{\prime}_x \sin \theta \right) \Psi(\vec{r}) +\cdots \label{eq:pbtrs}
\end{align}
Other possible terms (included in the ellipsis) are proportional to 
$\tau_z s_{x,y}$. Upon projection on the 0LL, Eq.~\eqref{eq:pbtrs} yields:
\begin{align}
\mathcal{P}V^{TRS}_{M}\mathcal{P} &= 
\int d\vec{r}\:  V_{\mathrm{pbiso}}(\vec{r}) \Psi^{\dag}_0(\vec{r})\Sigma_z 
\left(s^{\prime}_z \cos \theta  \right. \notag \\ 
 &\left.\qquad\qquad -  s^{\prime}_x \sin \theta \right) \Psi_0(\vec{r})+ \cdots 
\end{align}
Note that the second term on the right generates backscattering in the FM whereas the first one, which commutes with $s^{\prime}_z$, does not. In the CAF, both terms lead to backscattering because the spins are canted. However,  as pointed out above, this term is not a standard type of SOC as it requires sublattice exchange (parity)
symmetry to be broken. This may be caused by the substrate (BN), the tilted magnetic field, or the boundary conditions at the edge. However, estimating the strength of this term is beyond the scope of this work as this will require a detailed microscopic model of the system.  

\subsection{Non-time reversal symmetric scatterers}

Other possible potentials causing backscattering can break TRS. In particular, a potential of the form:
\begin{equation}
V^{NTRS}_{1M} =  \sum_{\alpha} h_{\alpha}(\vec{r}) s_{\alpha}, 
\end{equation}
which describes the exchange interaction with a locally magnetized scatterer (like a magnetized edge~\cite{Lado2014}),  gives a non-zero contribution 
to the backscattering potential the spins
of the edge electrons are canted and therefore do not need to be  completely flipped in order to backscatter.

 However, at large values of $B_{\parallel}$, for which canting is negligible,  $V^{NTRS}_{1M}$ will not produce backscattering in the  FM state. The reason is that  backscattering requires the potential to act both on the sublattice-valley and spin degree of freedom. In other words, the potential must flip between
(approximate) eigenstates of $\Sigma_x$ and $s^{\prime}_z$, meaning that the
orbital part must be proportional $\Sigma_z$ and $\Sigma_y = (\Sigma^{+} - \Sigma^{-})/2i$. 
 From the results in subsection~\ref{sec:project},  we see that $\Sigma_z$ arises from potentials proportional to $\sigma_0 \tau_z$
or $\sigma_z \tau_0$.  Note that $\tau_z$ is odd under TRS, and 
since flipping  the spin requires the potential be proportional  to $s_{\alpha}$, a term of the form $\tau_z s_{\alpha}$ is TRS.  The latter results in  the 
parity-symmetry
breaking SOC discussed above. On the other hand, $\sigma_z$ is TR even, 
and including the spin-part, leads to non-TRS potential of the form:
\begin{equation}
V^{NTRS}_{2M} = \sum_{\alpha=x,y,z} V^{\alpha}_{\mathrm{ntrs}}(\vec{r}) \sigma_{z} s_{\alpha}.
\end{equation}
Since $[V^{NTRS}_{2M}, \tau_z] = 0$, this potential
does not produce intervalley scattering. In addition, it also breaks sublattice exchange parity symmetry like $V^{TRS}_{M}$ from Eq.~\eqref{eq:pbtrs}.

Finally, potentials containing $\Sigma_y$ can arise from terms of the from
$\sigma_{\alpha}\tau_{\beta}$ with  $\alpha,\beta = x,y$, which
describe inter-valley scattering and are therefore are suppressed by a factor $\sim (a/l_B)^2$ as discussed above.

\section{Corrections to the Edge Conductance} \label{secIV}
In this section, we calculate the  
linear conductance of the edge channel in the presence of the scatterer. We focus on the \emph{doped} CAF, for which the leading sources of backscattering have been identified in subsection~\ref{sec:trs} and their strength can be estimated from their values in the zero-field Hamiltonian. The strength of the terms induced by
TRS- and/or sublattice parity symmetry-breaking terms is much more difficult to estimate because they are generated upon integrating high-energy states and will also depend on the microscopic details of the scatterer. In what follows, we shall use perturbation theory to lowest order in the scattering potential strength, $g$, assuming that the latter is weak, i.e. $g/\omega_c \ll 1$. 

As explained in Appendix~~\ref{app:contact} the correction to the channel conductance within linear response theory is given by: 
\begin{equation}
\delta G = \frac{dI_B}{dV}\Big|_{V=0}.
\end{equation}
where  $I_B$ is the backscattered current,
\begin{equation}
I_B =  -e \Big \langle\frac{d N_{R}}{dt} \Big \rangle,  
\end{equation}
$N_{R}$ being the total number of right moving electrons. Within linear response theory, the (steady-state) backscattered current is~\cite{Mahan}
\begin{eqnarray}
\begin{aligned}
I_B	=& (2  e) \: \mathrm{Im}\: \int dt\, e^{-ieVt} \: C_{AA^{\dag}}(t).
\end{aligned}
\end{eqnarray}
where the retarded correlation function is $C_{AA^{\dag}}(t) = -i \theta(t) \langle[A(t), A^{\dag}(0)]\rangle$ and  $A = g R_M  \psi^{\dag}_R(0) \psi_L(0)$.

 We first consider the non-interacting case for which $g_{RL}=0$. The
previous expression can be evaluated to yield:
\begin{multline}
I_B=-e (g R_M)^{2}\int_{-\infty}^{\infty}\frac{d\omega}{2\pi} \left[ G_{R}^{<}(eV+\omega)G_{L}^{>}(\omega) \right. \\
\left. -G_{R}^{>}(eV+\omega)G_{L}^{<}(\omega) \right]
\end{multline}
The non-interacting lesser and greater Green's functions are $G_{c}^{<}(\omega)=\frac{i}{v_{c}}n_{F}(\omega)$ and $G_{c}^{>}(\omega)=\frac{-i}{v_{c}}(1-n_{F}(\omega))$, where $n_{F}(\omega)$ is the Fermi-Dirac distribution and $c=R,L$. Hence, 

\begin{equation}
\delta G = \frac{(g R_M)^{2}}{v_{R}v_{L}}G_{0}\label{eq:nonint}
\end{equation}
For $v_{R} = v_L$, we recover the linear conductance of non-interacting 1D channel with an impurity.~\cite{Gia_QPin1D}

  However, note that the conductance of the non-interacting edge is independent of the temperature, in contradiction with the the experimental observations.~\cite{Young2014} This problem can be solved by accounting for 
interaction effects (i.e. $g_{RL}\neq 0$), which yield a temperature dependence of $\delta G$, in qualitative agreement with the experimental observation.~\cite{Young2014} Accounting for a non zero $g_{RL}$ is also necessary because interactions in 1D have a dramatic effect on the stability of fermionic quasi-particles 
.~\citep{Gia_QPin1D,RMP_Bosonization_MAC,AAA} 
Thus, we calculate the correction to the conductance using the bosonization method, which allows us to  treat interactions non-perturbatively
(see Appendices). The result of this 
calculation reads:
\begin{align}
\delta G(T) =& \frac{2 }{\pi T^{2-2K}} 
 \left(\frac{g R_{M}}{2\pi l_{B}}\right)^2 \left(\frac{4 \pi^{2} l_{B}^{2}}{v_{+}v_{-}}\right)^{K}  \sin(\pi K)
 \notag \\
& \times B(K,1-2K)\left[\psi(1-K)-\psi(K)\right] G_{0},
  \label{G}
\end{align}
where $K$ is the Luttinger parameter, which characterizes the strength of the interactions: $K = 1$ for $g_{RL} =0$ and $K < 1$ ($K > 1$) for repulsive (attractive) interactions.  The parameters  $v_{\pm}$ are the velocity of the eigenmodes  (see Appendix~\ref{sec: App_bos} for the definitions of $K$ and $v_{\pm}$); $B(x,y)$ is the beta function, $\psi(x)$ is the digamma function. Note that the temperature dependence
is a power-law, with an interaction dependent exponent,~\cite{Kane1992} which vanishes in the non interacting case (i.e. for $K = 1$), in agreement with Eq.~\eqref{eq:nonint}.

\begin{figure}[t]
\includegraphics[width=8cm]{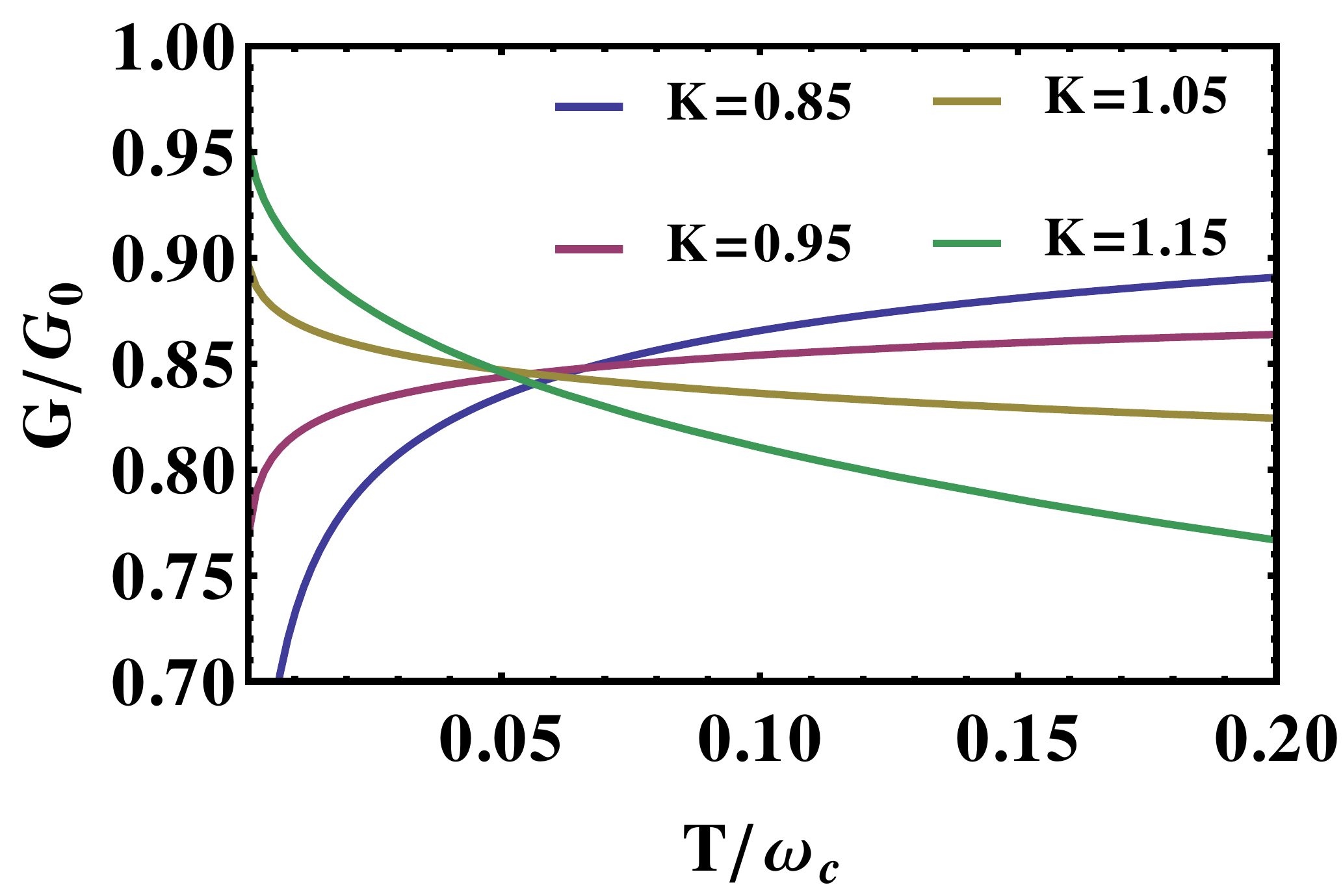}
\caption{ (Color online) Two-terminal conductance per edge $G = G_0 - \delta G$ ($G_0 = e^2/h$ is the quantized conductance) as a function of the temperature (in cyclotron frequency units), for different values of the Luttinger parameter $K$. This is plotted with the following parameters , $B_{\parallel}=45T$, $B_{\perp}=5T$, $R_M=l_{BS}=12nm$.}
\label{fig:G_vs_T}
\end{figure}

 In Fig.~\ref{fig:G_vs_T}, we have plotted the two-terminal conductance  per edge  against the temperature (in cyclotron frequency units) for different values of Luttinger parameter $K$. For $K>1$ (i.e. attractive interactions), the edge conductance approaches
the quantized value as $T\to 0$.  On the other hand, for $K<1$ (repulsive interactions), the 
deviation from perfect quantization increases
with decreasing $T$. For $\delta G \sim G_0$ linear response theory breaks down and we must rely on other methods, but this problem will not be analysed here. 

\begin{figure}[t]
\includegraphics[width=8cm]{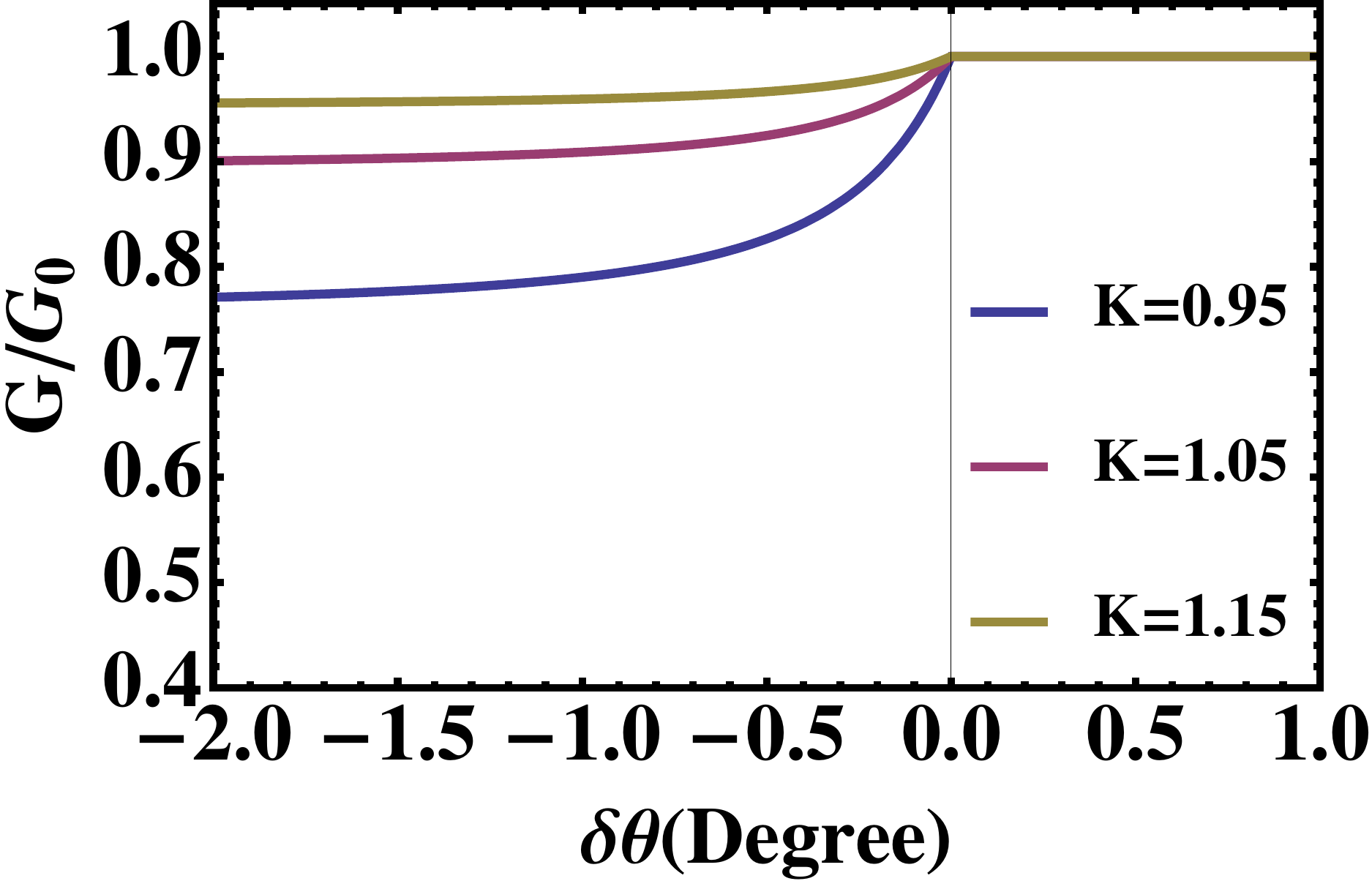}
\caption{ (Color online)  Two-terminal conductance per edge $G = G_0 - \delta G$ ($G_0 = e^2/h$ is the quantized conductance) as a function of the polar angle of the applied magnetic field $\delta\theta=\theta-\theta_{cr}$ (the perpendicular magnetic field $B_{\perp}$ is fixed), for different values of the Luttinger parameter $K$. 
The angle $\theta_{cr}= \cos^{-1}\left[ \epsilon_{z}/(2|u_{1}|)\right]$ is the critical angle for which the edge gap closes. Other calculation parameters used in this plot are chosen as follows $T=5$ K, $B_{\perp}=5$ T, $R_{M}=l_{B}=11.4$ nm.}
\label{fig:G_vs_theta}
\end{figure}

In Fig. \ref{fig:G_vs_theta}, the two-terminal conductance per edge is plotted against the deviation of the polar angle of the applied magnetic from the critical polar angle $\theta_{cr}$ at constant temperature, for different Luttinger parameter $K$.  

 Finally, let us briefly discuss the angular and temperature dependence of some of the parity-breaking SOC term
discused in Sec.~\ref{subsec:FM}. This term is expected to
yield a sizeable contribution  at large values of $B_{\parallel}$ in the FM state, when the canting~\cite{Murthy2014} of the edge channel spins is small.
 Upon projection onto the 0LL, Eq.~\eqref{eq:pbtrs} 
yields a term  $\propto \sin \theta$, where $\theta$ is the 
magnetic field polar angle. Hence, we obtain the following contribution of the conductance: 
\begin{equation}
\delta G_{\mathrm{pbiso}}(T,\theta) \propto \sin^2 \theta\,  T^{2K-2},
\end{equation} 
where $T$ is the absolute temperature and $K$ the Luttinger parameter. However, since this term is not present in the zero-field Hamiltonian, we cannot estimate its strength, although we expected it to be small. 


\section{Summary and Outlook}
\label{secVI}

Using a simple single-scatterer model, we have discussed the effects of potential and spin-orbit disorder on conduction through  helical edges recently observed in monolayer graphene under a    strong, tilted magnetic field.~\cite{Young2014} We have shown that the  deviation from perfect quantization
in the canted anti-ferromagnetic and ferromagnetic states
can be understood in terms of potential (i.e. scalar) disorder and to a lesser extent spin-orbit disorder of the
intrinsic type. Other types of disorder, such like 
sub-lattice exchange (parity) breaking spin-orbit coupling
and time-reversal symmetry-breaking spin-orbit coupling  terms are expected to yield smaller contributions.
Interestingly, unlike the situation encountered in 
semiconducting quantum wells,~\cite{Edge_Dynamics_QSH}
we find that Rashba spin-orbit coupling does not
lead to backscattering, at least to leading order. 

We have also investigated the temperature and magnetic field tilt angle dependence of a single-edge conductance assuming that the main scattering sources are the scalar potential and the (intrinsic) spin-orbit coupling induced by the scatterer. In addition,  the other scattering sources discussed in Sec. \ref{secIII} may be generated via virtual  transitions to different Landau levels, which
makes estimating their magnitude rather difficult. However, we expect them to be sub-dominant at large magnetic fields and therefore we  have neglected them in a first approximation. 

 We hope our work will shed light on the origin of the channel dissipation observed in the experiment. 
In particular, we have  found that disorder-induced backscattering in combination with electron-electron interactions leads to a power-law temperature dependence of the conductance which may be observed experimentally. In this regard, we note that, even in the ferromagnet, for which the backscattering effects are milder, the presence of impurities, combined with repulsive interactions can lead to an insulating-like behaviour of the edge conductance. 

The model studied here can be fairly easily generalized to disorder potentials with more complicated  spatial dependence.  Such a study will not be pursued here. However, we can anticipate~\cite{Gia_QPin1D} it will yield to a different exponent in the temperature dependence of the conductance.A thorough understanding of the origin of dissipation in the helical edges of graphene under a strong, magnetic field will require a more detailed microscopic model of the edge of the device. Such model should account of the existence of multiple  channels~\cite{Gusynin2008} and also the coupling with the gapless bulk (Goldstone) modes.~\cite{Murthy2014}

\acknowledgments

C.H thanks V.M. Pereira and M. Milletari for discussions and the Graphene Research Center of the National University Singapore for support during the early stages of this work. We also gratefully acknowledge discussions with P. Jarillo-Herrero. 
C.H's work was supported in part by the Singapore National Research
Foundation grant No.~NRFF2012-02. MAC's work is supported by the Ministry of Science and Technology (MoST) of Taiwan. 
  
\appendix

\section{Bosonization}
\label{sec: App_bos}

In this and the following Appendices, we collect some of the most important results of bosonization technique~\cite{Gia_QPin1D} applied to a one-dimensional (1D) channel with two different carrier velocities, which is relevant to the situation studied in the main text. 

The Hamiltonian of the 1D edge can be expressed in terms of the Hamiltonian density, $\mathcal{H}_{\mathrm{edge}}(x)$, as $H_{\mathrm{edge}}=\int dx\: \mathcal{H}_{\mathrm{edge}}(x)$, and $\mathcal{H}_{\mathrm{edge}}(x)$ can be fully expressed in terms of the fermion chiral
density operators,
\begin{equation}
\rho_{c}(x) = \, :\psi^{\dag}_c(x)\psi_c(x): \,
\end{equation}
where $c = R, L$ and $:\ldots:$ stands for normal order. 
The chiral densities obey the Kac-Moody algebra:~\cite{Gia_QPin1D}
\begin{equation}
\left[\rho_{c}(x), \rho_{c^{\prime}}(x^{\prime})\right] = \frac{\delta_{c,c^{\prime}}s_c }{2\pi i}
\partial_{x}\delta(x-x^{\prime}).
\end{equation}
where $s_{R} = -s_L = +1$. Thus, the 
Hamiltonian density can be written as:~\cite{Gia_QPin1D}
\begin{align}
\mathcal{H}_{\mathrm{edge}} = \pi :\left[ v_{R} \rho^2_R + v_{L} \rho^2_L + 2 g_{RL}
\rho_R \rho_L \right]:\, , 
\end{align}
where $v_{R}$ ($v_{L}$) is the Fermi velocity of the right (left) channel and $g_{RL}$ is an effective short-range electron-electron interaction. 
It is also useful to introduce the bosonic fields $\phi_{c=R,L}(x)$, 
which are defined by  
\begin{equation}
\rho_{c}(x) =  \frac{1}{2\pi} \partial_x \phi_{c}(x).
\end{equation}

The Hamiltonian density can be brought to diagonal form:
\begin{equation}
\mathcal{H}_{\mathrm{edge}}(x) =  \frac{\pi}{K} :\left[ v_{+} \rho^2_{+}(x) + v_{-} \rho^2_{-}(x) \right]:\,, \label{eq:hamdens}
\end{equation}
by means of the following linear transformation:
\begin{equation}
\left( \begin{array}{c}
\rho_{+}(x) \\
\rho_{-}(x)
\end{array} \right)
= \frac{1}{2} 
\begin{pmatrix}
 1 + K &  1 - K  \\
1 -K  &  1 + K
\end{pmatrix}
\left( \begin{array}{c}
\rho_{R}(x) \\
\rho_{L}(x)
\end{array} \right),
\label{eq: Bogoliubov_rot}
\end{equation}
where
\begin{equation}
K = \sqrt{\frac{v_R + v_L - 2g_{RL}}{ v_R + v_L + 2g_{RL} }}.
\end{equation}
and
\begin{equation}
v_{\pm} =  \pm \frac{ (v_R - v_L)}{2} + \sqrt{\left( \frac{v_R +v_L}{2}\right)^2- g_{RL}^2}. 
\end{equation}
Note that the transformation in  Eq.~\eqref{eq: Bogoliubov_rot} has the following properties:
\begin{align}
\rho_{+}(x) + \rho_{-}(x) &= \rho_{R}(x) + \rho_L(x),\\
\rho_{+}(x) - \rho_{-}(x) &= K \left[ \rho_{R}(x) - \rho_{L}(x) \right] \label{eq:diffrho}.
\end{align}
The new chiral fields obey a modified Kac-Moody algebra:
\begin{equation}
\left[\rho_{\alpha}(x), \rho_{\beta}(x^{\prime})\right]
= \frac{s_{\alpha}  K \delta_{\alpha\beta}}{2\pi i}\partial_{x}\delta(x-x^{\prime}),
\end{equation}
where $\alpha,\beta = \pm$ and $s_{\pm} = \pm 1$. Using these commutation relations, the chiral densities are found to obey the following equations of motion:
\begin{equation}
\left(\partial_{t} \pm v_{\pm} \partial_{x} \right) \rho_{\pm}(x,t) = 0. \label{eq:ecmot}
\end{equation}
In the stationary state (i.e. for $\partial_{t}\rho_{\pm} = 0$), when coupled to two different (chiral) chemical potentials, the chiral densities must minimize  
$\mathcal{H}_{\mathrm{edge}}(x) - \mu_{+} \rho_{+}(x) - \mu_{-}\rho_{-}(x)$, 
which is possible (upon completing the square) if 
we choose:
\begin{equation}
\rho_{\pm} = \langle \rho_{\pm}(x)\rangle   = \frac{K\mu_{\pm}}{2\pi v_{\pm}}.\label{eq:cdens}
\end{equation}
Hence the stationary current is given by
$I = e (v_{+}\rho_{+} - v_{-}\rho_{-}) = e K (\mu_{+}-\mu_{-})/(2\pi)$. Note that
the chemical potentials, $\mu_{\pm}$ coupling to the densities 
$\rho_{\pm}$ are physically different from the chemical potentials, $\mu_{R/L}$, coupling to the electrons.
The latter couple to $N_{R/L} = \int dx \rho_{R/L}(x)$
and are equal to the chemical potentials of the source and
drain lead reservoirs, $e V_{S/D}$, prespectively. 
However, it is possible to relate both sets
of chemical potentials linearly,~\cite{Pham} so that 
the voltage bias $V = V_{S}-V_{D} = K (\mu_{+} - \mu_{-})/e$.  Hence, the quantized conductance for a clean 1D 
edge channel can be recovered:~\cite{maslov1995landauer,kawabata2007electron,Pham}
\begin{equation}
G = \frac{dI}{dV} = \frac{e^2}{2\pi}.
\end{equation}
which is the value of the quantized conductance for a single channel in $\hbar = 1$ units. The quantized conductance of 
a perfect non-chiral Tomonaga-Luttinger liquid is the result of the finite contact resistance.
 
\section{Accounting for the Contact Resistance}
\label{app:contact}
 In this Appendix, we shall follow the approach of Pham and coworkers~\cite{Pham} to account for the contact resistence of the 1D interacting edge channel in the presence of a scatterer located at $x = 0$.  As it has been discussed in the main text, the Hamiltonian for such system is 
$H = H_{\mathrm{edge}} + V_{M} = \int dx \, \mathcal{H}$,
where the Hamiltonian density is given by: 
\begin{equation}
\mathcal{H} = \mathcal{H}_{\mathrm{edge}} + g R_{M}\left[ \psi^{\dag}_{R}(x)\psi_L(x) +  \mathrm{h.c.}  \right]\delta(x)
\end{equation}
We are interested in computing the two terminal conductance, which is defined from the current passing through the device, $I_{2t}$, as follows:
\begin{equation}
G = \frac{dI_{2t}}{dV}\Big|_{V=0},
\end{equation}
where $V = V_{S}-V_{D}$ is the voltage bias between the source and drain leads. The current $I_{2t}$ is given by
the following equations (essentially Ohm's law):~\cite{Pham}
\begin{eqnarray}
V_{S}-\frac{\bar{\mu}^{<}}{e}=R_{S}I_{2t}, \label{eq:ohm1}\\
\frac{\bar{\mu}^{>}}{e}-eV_{D}=R_{D}I_{2t},\label{eq:ohm2}
\end{eqnarray}
where
\begin{equation}
\bar{\mu}^{\lessgtr}=\frac{\mu_{+}^{\lessgtr}+
\mu_{-}^{\lessgtr}}{2}.
\end{equation}
are the mean chemical potentials of the edge  to the right ($\bar{\mu}^{>}$) and left ($\bar{\mu}^{<}$) of the impurity, respectively; $R_{S},R_{D}$ ($ V_{S},V_{D}$) are the the source and drain contact resistances (voltages), respectively.

 In the presence of the impurity, the equations of motion
for $\rho_{\pm}$ (cf. Eq.~\ref{eq:ecmot}) are modified and
can be written as follows:
\begin{equation}
\left( \partial_{t} \pm v_{\pm} \partial_{x}\right) \rho_{\pm}(x,t) = J_{\pm}(t) \delta(x),\label{eq:ecmot2}
\end{equation}
where the operator $J_{\pm} = i \int dx\, \left[V_M, \rho_{\pm}(x) \right] = i \left[ V_{M}, N_{\pm} \right]$, with $N_{\pm} = \int dx\: \rho_{\pm}(x)$. However, note that, away from the scatterer (i.e. $x\neq 0$), 
the equations of motion reduce to Eq.~\eqref{eq:ecmot}. Thus, in the steady state, the chemical potentials $\mu^{\lessgtr}_{\pm}$ can be related to the chiral densities by means of Eq.~\eqref{eq:cdens}, that is, 
\begin{eqnarray}
\rho_{\pm}^{\lessgtr}=\frac{K}{2\pi v_{\pm}}\mu_{\pm}^{\lessgtr}.
\end{eqnarray}
where $\rho^{\lessgtr}_{\pm}$ are the chiral densities to the right and left of the scatterer.  Adding equations \eqref{eq:ohm1} and \eqref{eq:ohm2} yields:
\begin{equation}
I_{2t}=\frac{V_{S}-V_{D}}{R_{S}+R_{D}}-\frac{R_0}{R_{S}+R_{D}}I_{B},\label{eq:ohm}
\end{equation}
where we have introduced $R_0 = G^{-1}_0 = 2\pi/e^2$ and  
\begin{align}
I_{B}= -\frac{e}{2K}\left[ v_{+}(\rho^{>}_{+}-\rho^{<}_{+}) + v_{-}(\rho^{>}_{-}-\rho^{<}_{-}) \right]  
\end{align}
which, as shown below, is the back-scattered current. Note that our definition of  $I_{B}$ differs from the definition in Ref.~\onlinecite{Pham} by a factor of $K^{-1}$; $I_B$ can be obtained upon integraging over an infinitesimal interval
around $x =0$ the above equations of motion~ Eq.~\eqref{eq:ecmot2} and taking the expectation value, which yields:
\begin{equation}
v_{\pm} \left(\rho^{>}_{\pm} - \rho^{<}_{\pm}\right) 
= \pm   \langle J_{\pm} \rangle.
\end{equation}
Hence, 
\begin{align}
I_{B} &= -\frac{e}{2K} \langle \left(J_{+} - J_{-} \right) \rangle = -\frac{i e}{2K} \langle \left[V_{M}, N_{+}-N_{-} \right]  \rangle \notag \\ 
&= -\frac{i e}{2} \langle \left[ V_{M}, N_{R}-N_{L} \right] \rangle \notag\\ 
&= -\frac{i e}{2} \langle \left[ H, N_{R}-N_{L} \right] \rangle.
\end{align}
In deriving the last expression, we have integrated Eq.~\eqref{eq:diffrho} over $x$ and used that $[H_{\mathrm{edge}}, N_{R/L}] = 0$. Next we use $N_{R}-N_{L} = 2N_{R} -N$ and $[H, N]=0$, where $N = N_{R}+N_{L}$, which leads to
\begin{equation} 
\hat{I}_{B} = (-ie) \left[ H, N_{R}\right] =  -e \Big \langle \frac{dN_{R}}{dt} \Big\rangle 
\end{equation}
Hence, assuming a symmetric  situation where $R_{S} = R_{D} = \pi/e^2$, it follows that
\begin{equation}
I_{2t} = G_{0} V - I_B, \label{eq:current}
\end{equation}
where $G_{0}= e^2/2\pi$ is the clean channel conductance. After derivation with respect to the voltage bias, Eq.~\eqref{eq:current} for $V = 0$  becomes:
\begin{equation}
G = G_0 - \delta G,
\end{equation}
$\delta G = (dI_B/dV)_{V=0}$ being the conductance across the impurity. The latter is computed  in the following Appendix.

\section{Linear Response for the Conductance} \label{app:lin}

In this appendix, we review the calculation of the linear conductance of a 1D channel in the presence of an impurity, $\delta G$, using the Kubo formula. The result is perturbative in the strength of the impurity potential and requires $\delta G \ll G_0$, which, as discussed in the main text will break down as the absolute temperature $T\to 0$ for repulsive interactions ($K < 1$).  We begin by introducing the backscattering current operator:  
\begin{equation}
\label{eq:bs_current}
\hat{I}_B= -e  \frac{dN_{R}}{dt} = (-i e) [H,N_{R}] =  (-i e) \left[ A - A^{\dag}\right],
\end{equation}
where $H$ is given by  Eq.~\eqref{eqn:H} and the operator $A = g R_{M} \psi^{\dag}_{R}(0) \psi_{L}(0)$. Within linear response theory, the steady state current in response to a voltage bias $V$ 
is (see e.g. Ref.~\onlinecite{Mahan}, page 561 and ff.):
\begin{equation}
I	= 2 e \:\mathrm{Im}  \int_{-\infty}^{+\infty} 
dt \:  e^{-ieVt} \, C^{R}_{AA^{\dag}}(t)
\end{equation}
where the correlation function, $C^{R}_{AA^{\dag}}(t) = -i \theta(t) \langle \left[A(t), A^{\dag}(0)  \right] \rangle$
where $\theta(t)$ is the Heaviside step function. In order
to evaluate the above correlation function, we rely
on bosonization, which uses the following representation of 
the Fermi fields in terms of the bosonic fields $\phi_{R/L}(x)$ introduced in Appendix~\ref{sec: App_bos}:
\begin{eqnarray}
\psi_{R}(x)=\frac{F_{R}}{\sqrt{2\pi l_{B}}}e^{i\phi_{R}(x)} , \\
\psi_{L}(x)=\frac{F_{L}}{\sqrt{2\pi l_{B}}}e^{-i\phi_{L}(x)} ,
\end{eqnarray}
where $F_{c} = F^{\dag}_{c}$ ($c = R, L$) are the Klein factors, 
$\{F_{c}, F_{c^{\prime}}\} = \delta_{cc^{\prime}}$, required to ensure the anti-commutativity of the right and left moving Fermi fields. 

 In the presence of interactions, the bosonic fields $\phi_{R/L}(x)$ do not describe the eigenmodes of the clean interacting 1D edge. However,  we can introduce a pair of bosonic fields $\phi_{\pm}(x)$
reated to the chiral densities $\rho_{\pm}$ that diagonalize $H_{\mathrm{edge}}$: 
\begin{equation}
\rho_{\pm}(x) = \frac{\sqrt{K}}{2\pi}\partial_{x}\phi_{\pm}(x).
\end{equation}
The finite temperature correlation function of the new bosonic fields (for $|t|\gg l_{B}/v_{\pm}$) at $x=0$ reads:~\cite{Gia_QPin1D}
\begin{equation}
\langle\phi_{\pm}(0,t)\phi_{\pm}(0,0)\rangle=-\ln\left[\left(\frac{2v_{\pm} \beta }{\pi l_B}\right) \sin\left( \frac{i\pi t}{\beta} \right) \right].
\label{eq: Bosonic green func}
\end{equation}
Using  Eq. \eqref{eq: Bosonic green func}, the correlator $C_{AA^{\dag}}(t)$ can be evaluated and thus we arrive at the following expression for the current:
 \begin{equation}
 I=2e(gR_{M})^{2}C \, \mathrm{Re} \int_{-\infty}^{0}dt'e^{ieVt'}\{f(-t)-f(t)\},
 \end{equation}
where
\begin{equation}
f(t)=\left[\sin\frac{\pi}{\beta}\left(it+\frac{l_{B}}{v_{+}}\right)\sin\frac{\pi}{\beta}\left(it+\frac{l_{B}}{v_{-}}\right)\right]^{-K},
\end{equation}
and the prefactor
\begin{equation}
C= \left(\frac{1}{2\pi l_{B}}\right)^2\left[\frac{\pi^{2}l_{B}^{2}}{\beta^{2}v_{+}v_{-}}\right]^{K}.
\end{equation}
After integration, we obtain:
\begin{align}
I_B = &  \frac{4 C e (g R_M)^2 \beta}{2^{1-2K}\pi}  \sin (\pi K) \: 
\mathrm{Im}   \: B\left(K+\frac{ieV\beta}{2\pi},1-2K\right)\notag \\
\end{align}
where $B(n,m)$ is the beta function. From this expression, after
derivation with respect to $V$, at $V = 0$,  Eq.~\eqref{G} follows.

\bibliographystyle{apsrev4-1}
\bibliography{QSH_graphene}

\end{document}